\documentclass[fleqn,10pt]{wlscirep}
\usepackage[utf8]{inputenc}
\usepackage[T1]{fontenc}
\usepackage{float}
\usepackage{newunicodechar}
\newunicodechar{ }{~}
\usepackage{multirow} 
\usepackage{soul}
\usepackage{xcolor}
\definecolor{mygreen}{RGB}{34,139,34}

\title{Polarity-Resolved Far-Side Magnetograms Based on Helioseismic Measurements}

\author[1,*]{Amr Hamada}
\author[1]{Kiran Jain}
\author[2]{Hanna Strecker}
\author[3]{Charles Lindsey}
\author[2]{David Orozco Su{\'a}rez}
\affil[1]{National Solar Observatory, Boulder, CO 80303, USA}
\affil[2]{Instituto de Astrofísica de Andalucía, CSIC, Granada 18008, Spain}
 \affil[3]{North West Research Associates, Boulder, CO 80301, USA}

\affil[*]{ahamada@nso.edu}


\begin{abstract}
Understanding and monitoring solar active regions is essential for operational space-weather forecasting and better solar dynamo modeling. This requires comprehensive 360$^\circ$ observations of the Sun. While space-weather forecasting has long relied successfully on high-quality observations of the Earth-facing hemisphere, a critical gap in global magnetic context remains due to the lack of direct, continuous magnetic field measurements of far-side active regions, specifically magnetic field strength, polarity configurations, and related parameters. We present a methodology for inferring magnetic field distributions of active regions in helioseismic maps of the far hemisphere. The crux of the analysis is the ability to realistically surmise the signs of the magnetic polarities of opposing components of a helioseismic signature. We present a method for stable, continuous polarity assignment of large-scale magnetic structures, derived from substructures that helioseismic signatures reliably resolve in strong active regions--particularly those that become space-weather hazards as solar rotation brings them into Earth’s view. Polarity boundaries are identified by analyzing the bi-modal longitudinal variance profile of the seismic signal within each region, after which Hale’s polarity rule is applied to establish east–west ordering consistent with the solar cycle. The method yields polarity-resolved far-side magnetograms that are suitable for integration with near-side observations, enabling the construction of full-Sun magnetic boundary conditions for coronal and solar-wind modeling, and providing a critical step toward improved heliospheric simulations and operational forecasting.

\end{abstract}

\begin{document}

\flushbottom
\maketitle
%
%
\thispagestyle{empty}

\section*{Introduction}

Solar active regions (ARs) are areas of strong magnetic field, and are the drivers of space weather. Their emergence, evolution, and decay are frequently accompanied by flares and coronal mass ejections (CMEs) that disturb the heliosphere and impact geospace systems. A comprehensive understanding of space weather and its predictive capabilities requires concurrent monitoring of the Sun’s Earth-facing and far-side hemispheres\cite{Jain_2023}. While Earth-side magnetography provides continuous, detailed coverage of the visible disk, characterizing ARs on the solar far-side remains challenging due to the lack of magnetic measurements. Although flux transport models -- such as the Advective Flux Transport Model \cite{2024ApJ...968..114U} -- are commonly used to estimate magnetic flux from regions that have rotated beyond the west limb, forecast accuracy near the east limb remains significantly limited. Far-side observations are particularly vital when ARs evolve rapidly or when new, strong regions emerge on the unseen hemisphere.

Since the turn of the century, helioseismic holography and related phase-correlation methods have given us a valuable proxy by revealing the existence of newly emerged far-side ARs from perturbations of acoustic waves recorded on the near-side (see C. Lindsey \& D. Braun \cite{Lindsey2017} for review). Applied to Dopplergrams from the Global Oscillation Network Group\cite{Harvey1996,Jain2021} (GONG) and the Helioseismic and Magnetic Imager\cite{HMI2012} onboard Solar Dynamics Observatory spacecraft (SDO/HMI), these techniques routinely detect large to moderate ARs on the far hemisphere that feed daily operational far-side synoptic products\cite{igh2010}. Direct validation of these helioseismic far-side detections through comparisons with Solar TErrestrial RElations Observatory/Extreme Ultraviolet Imager (STEREO/EUVI) observations has demonstrated that the resulting far-side AR images are highly reliable and accurately track the emergence and evolution of ARs \cite{Liewer_2017,Zhao_2019}. To advance both scientific interpretation and model assimilation, however, the ARs made apparent by helioseismology need to be realistically resolved into the opposing magnetic polarities of their components, a crucial property to which helioseismic signatures, by themselves, are insensitive.

Several studies have integrated far-side maps into modeling frameworks to assess the influence of far-side ARs on forecast accuracy. Case studies involving solar wind and coronal hole backcasting, where far-side data were assimilated, have shown encouraging results \cite{Arge_2013}. These efforts utilized synoptic magnetic flux maps from the Air Force Data Assimilative Photospheric Flux Transport (ADAPT) model as input to the Wang–Sheeley–Arge (WSA) model, which requires a global, instantaneous magnetic field to define inner boundary conditions for solar wind prediction. At present, ADAPT maps are constructed solely from line-of-sight full-disk magnetograms and exclude far-side ARs \cite{Arge2010}. They demonstrated that integrating far-side ARs, particularly those near the solar limbs, can significantly improve WSA model predictions by refining estimates of the global magnetic flux distribution. In a separate study, the short-term forecast of UV irradiance, including Ly$\alpha$, was significantly improved by incorporating farside seismic information \cite{2009AdSpR..44..457F}.

Early systematic calibrations leveraged near-side “bookend” measurements around far-side transits. I. Gonz{\'a}lez Hern{\'a}ndez et al.\cite{Gonzalez_2007} compared GONG-based far-side phase signatures with photospheric magnetic and sunspot measurements before and after transit, establishing significant correlations between the seismic phase signal and both unsigned line-of-sight (LoS) magnetic field and spot area. Building on this, G. A. MacDonald  et al.\cite{MacDonald_2015} combined near-side GONG helioseismology with ADAPT synchronic magnetic maps, confirming morphological consistency (e.g., area, tilt) and showing that a logarithmic relation between phase shift and unsigned LoS magnetic field holds for co-temporal measurements. Their analysis employed a different set of modes than those used in far-side helioseismology.

More recently, several studies have sought to quantitatively calibrate the helioseismic phase signal to magnetic field. R. Chen et al.\cite{Chen_2022} demonstrated an empirical linkage between helioseismic phase shifts and far-side unsigned magnetic flux by combining SDO/HMI time-distance maps with STEREO/EUVI-based magnetic proxies. Using Solar Orbiter\cite{SO,2020_Muller}/Polarimetric and Helioseismic Imager\cite{PHI} (SO/PHI)  magnetograms with HMI-derived seismic maps of six far-side ARs, D. Yang et al.\cite{Yang_2023} performed the first direct validation of far-side helioseismic holography. They established a distinctly monotonic relationship between the mean helioseismic signature and the unsigned line-of-sight magnetic field, supporting the use of phase-shift magnitude as a quantitative proxy for unsigned LoS magnetic field intensities.

Building on these advances, E. G. Broock et al.\cite{Broock_2024} developed the FarNet-II deep-learning architecture to infer far-side magnetograms, including polarity, from seismic data, achieving quantitative agreement with SO/PHI observations. To advance both scientific interpretation and model assimilation, however, the helioseismic signal must be quantitatively linked to magnetic-field properties and validated against direct magnetograms when available. All of these results support the calibration methodology adopted here and align with our FArSide Trained Active Region Recognition (FASTARR) pipeline\cite{Hamada_2025}. Despite this progress, a key gap persists, as the polarity-resolved quantities on the far side—including magnetic field, polarity assignment, tilt angle, and related parameters, are not routinely available, despite recent efforts exploring polarity inference from helioseismic diagnostics \cite{Yang_2024}.  

This study addresses a critical gap by leveraging the coherence between helioseismic phase signals and magnetic field properties to reconstruct far-side magnetic information compatible with data-driven and machine-learning frameworks. The proposed methodology establishes a novel framework that is structurally and conceptually distinct from previous approaches. In an earlier study, Arge et al. devised a magnetic flux distribution over the region inferred from the helioseismic signature first by bisecting it along the longitude that divides it into equal areas\cite{Arge_2013}. The opposing fluxes, inferred from the strength of the helioseismic signature, are then distributed uniformly over their respective regions, their polarities prescribed by the Hale polarity law. While these approaches demonstrate the feasibility of polarity inference on the far-side, the present work introduces a structurally distinct framework that emphasizes automated region selection, quantitative phase–magnetic calibration, and temporally stable polarity assignment.

As this configuration appears to perform well for modeling the global magnetic field tens to hundreds of megameters into the solar corona, it relies on an unrealistically thin current sheet separating the two halves. Front-side magnetograms suggest that such a structure is generally unrealistic for analysts focused on the magnetohydrostatic (MHS) conditions required to sustain it -- and even less plausible under magnetohydrodynamic (MHD) processes, which would likely disrupt it rapidly and violently.

The algorithm we exercise here undertakes to devise a more continuous distribution of magnetic field and its polarity that satisfies (1) an empirical calibration of the non-linear relationship of the helioseismic signature with the magnetic pressure, (2) a constraint that the estimated magnetic field within the region identified by FASTARR remains consistent with the physical expectation of a nearly balanced magnetic configuration, and (3) that the Hale polarity be satisfied. This approach utilizes longitudinal profiles of the helioseismic signature integrated over latitude, and latitudinal profiles of the same integrated over longitude to devise a fully continuous 2-dimensional profile that meets the three criteria above, hence devoid of precipitous current sheets. This accommodation that follows, then, is intended to exploit our helioseismic maps to give us more MHD-realistic polarity-assigned magnetic reconstructions of the solar far side, enabling daily full-Sun magnetograms for assimilation into space weather forecasting models, such as coronal and solar-wind models.\\

\section*{Dataset}
\label{dataset}
This study employs a coordinated set of far-side measurements to evaluate the relationship between the helioseismic phase-shifts and corresponding magnetic structures. The dataset spans within three consecutive years (2022--2024), starting in April to October/September, and is composed of three primary components: (1) GONG phase-shift maps, (2) AR masks generated by the machine-learning (ML) FASTARR model \cite{Hamada_2025}, and (3) Level-2 Carrington-projected LoS magnetograms from SO/PHI. These magnetograms are represented as B throughout the text. The full summary of these datasets is provided in Table~\ref{table1}, detailing the source, and time range of each data product. 

All the phase-shift maps are temporally averaged following the same approach adopted by A. Hamada et al.\cite{Hamada_2024}. In brief, we begin by computing phase-shift maps from approximately 24 hours of 1-minute cadence GONG Dopplergrams using helioseismic holography. To ensure the integrity of the input Dopplergrams, we apply a machine-learning based filter\cite{Jain_2025} to eliminate erroneous input Dopplergrams that could introduce spurious noise into the resulting maps. Each reference map is then temporally averaged using a sliding window that combines each map with its three immediate predecessors, corresponding to an 18-hour averaging window in map cadence, while preserving a 6-hour sampling. This method devises temporal smoothing and enhances the signal-to-noise ratio, while preserving the large scale structural evolution of far-side features. Every map is marked with a Duty Cycle (DC), a number from 0 to 1, which represents the fraction of the 1440 minutes time series for which a good observation was acquired. To maintain data quality consistency across the averaged maps, we applied a DC threshold by excluding from the averaging process all individual phase maps computed from datasets with a DC below 0.8. The resulting averaged helioseismic phase-shift maps are then used to identify and characterize far-side AR acoustic signatures. 

The AR masks are generated by the ML–based FASTARR model at the same nominal cadence, providing consistent spatial boundaries for each far-side AR and enabling subsequent statistical analysis of morphology and phase-related properties. To validate the magnetic nature of the far-side detected regions, we utilize SO/PHI magnetograms of the Sun's far side co-spatial and concurrent with the helioseismic signature thereof. These magnetograms offer a unique LoS perspective of the Sun's far side and are temporally synchronized with the helioseismic GONG's measurements during the selected times. This synchronic acquisition between the helioseismic and magnetic dataset represents a major methodological advance over the previous studies, which had to rely on temporally offset datasets, by minimizing AR evolution during the temporal delay between far-side transit and direct observations. The elimination of this delay greatly reinforces the correlation between the seismic signatures and underlying magnetic structures. The SO/PHI synoptic magnetic maps were produced by first transforming the helioprojective image coordinates into heliographic coordinates using the individual Solar Orbiter ephemeris information. Each Level-2 line-of-sight magnetogram was then reprojected onto a regular Carrington longitude–latitude grid using a plate-carree (CAR) projected world coordinate system with uniform pixel spacing. This reprojection performs the coordinate transformation and resampling simultaneously, yielding maps with a 1-pixel sampling that matches the spatial sampling of the GONG phase-shift maps.

\begin{table}[t]
\begin{center}
\caption{Datasets used for far-side solar analysis. AR masks were generated using the FASTARR model trained on GONG phase-shift maps.}
\label{table1}
\begin{tabular}{lcccc}
\hline
\multicolumn{1}{c}{Far-Side Dataset} & {Source / Instrument} & {Observation Times [UT]} & {Start} & {End} \\
\hline
\multirow{3}{*}{GONG phase-shift Maps} & \multirow{3}{*}{GONG Network} & \multirow{3}{*}{00:00, 06:00, 12:00, 18:00} & 2022 April 08 & 2022 September 20 \\
 &  &  & 2023 April 26 & 2023 October 01 \\
 &  &  & 2024 April 16 & 2024 September 26 \\
\cline{1-5}
\multirow{3}{*}{AR Masks} & \multirow{3}{*}{FASTARR ML algorithm} & \multirow{3}{*}{00:00, 06:00, 12:00, 18:00} & 2022 April 08 & 2022 September 20 \\
 &  &  & 2023 April 26 & 2023 October 01 \\
 &  &  & 2024 April 16 & 2024 September 26 \\
\cline{1-5}
\multirow{3}{*}{LOS Magnetograms} & \multirow{3}{*}{Solar Orbiter / PHI} & \multirow{3}{*}{1--8 maps/day (irregular)} & 2022 April 08 & 2022 September 20 \\
 &  &  & 2023 April 26 & 2023 October 01\\
 &  &  & 2024 April 16 & 2024 September 26 \\
\hline
\end{tabular}
\end{center}
\end{table}

We adopt a four-times daily cadence at 00:00, 06:00, 12:00 and 18:00 UT from the GONG phase-shift maps as a temporal reference for tracking the evolution of far-side ARs. This regular sampling enables temporally resolved analysis of AR development. In contrast, the cadence of SO/PHI magnetograms varies significantly, with daily image counts ranging from one to more than eight, and no consistent temporal pattern. To facilitate systematic comparison, we implemented an automated procedure that, for each day in the selected interval, selects the SO/PHI magnetogram closest to that reference time (00:00, 06:00, 12:00, and 18:00 UT). A magnetogram is retained only if the time difference from the target reference is less than or equal to 06 hours. The resulting set of timestamps and filenames forms the basis for aligning GONG-derived helioseismic AR detections with available SO/PHI magnetic field observations. The temporal coverage listed in Table~\ref{table1} corresponds to the far-side visibility constraints described in the following sub-section.

\subsection*{Far-Side SO/PHI Visibility}
\label{farside_geometry}
To evaluate the correspondence of the far-side helioseismic AR detections with SO/PHI magnetograms, we first defined the periods during which SO/PHI had favorable viewing geometry of the solar far-side. As there is no universally defined criterion for such periods, we adopted a geometric proxy based on the Solar Orbiter-Sun–Earth (SO–S–E) separation angle. Using Solar Orbiter ephemerides, we computed the SO–S–E separation angle for each day. The key selection criterion was based on the longitudinal separation angle ($\theta$) between Solar Orbiter and Earth in Carrington coordinates, normalized to the range $[-180^\circ, +180^\circ]$. At $\theta = 0^\circ$, Solar Orbiter is aligned with Earth, having the same solar hemispheric view (front-side). When the angle $\theta = \pm180^\circ$, Solar Orbiter is on the opposite side of the Sun, viewing the full far-side hemisphere. The normalized (SO–S–E) separation is defined as
\begin{equation}
x \equiv \left|\frac{\theta}{180} \right|, {\rm ~with} ~\theta \in [-180^\circ, ~+180^\circ], {\rm ~whereby} ~x \in [0, 1].
\end{equation}

All the positive angular values indicate that Solar Orbiter is ahead of Earth in Stonyhurst longitude, while the negative values indicate it is behind. This approach provides a geometrically consistent and observationally constrained dataset, while accounting for the annual variability, occasional data gaps, and the asymmetric temporal coverage of certain separation angles around superior conjunction by the spacecraft's orbit.

\begin{figure}[t]
    \centering
    \includegraphics[width=0.8\textwidth]{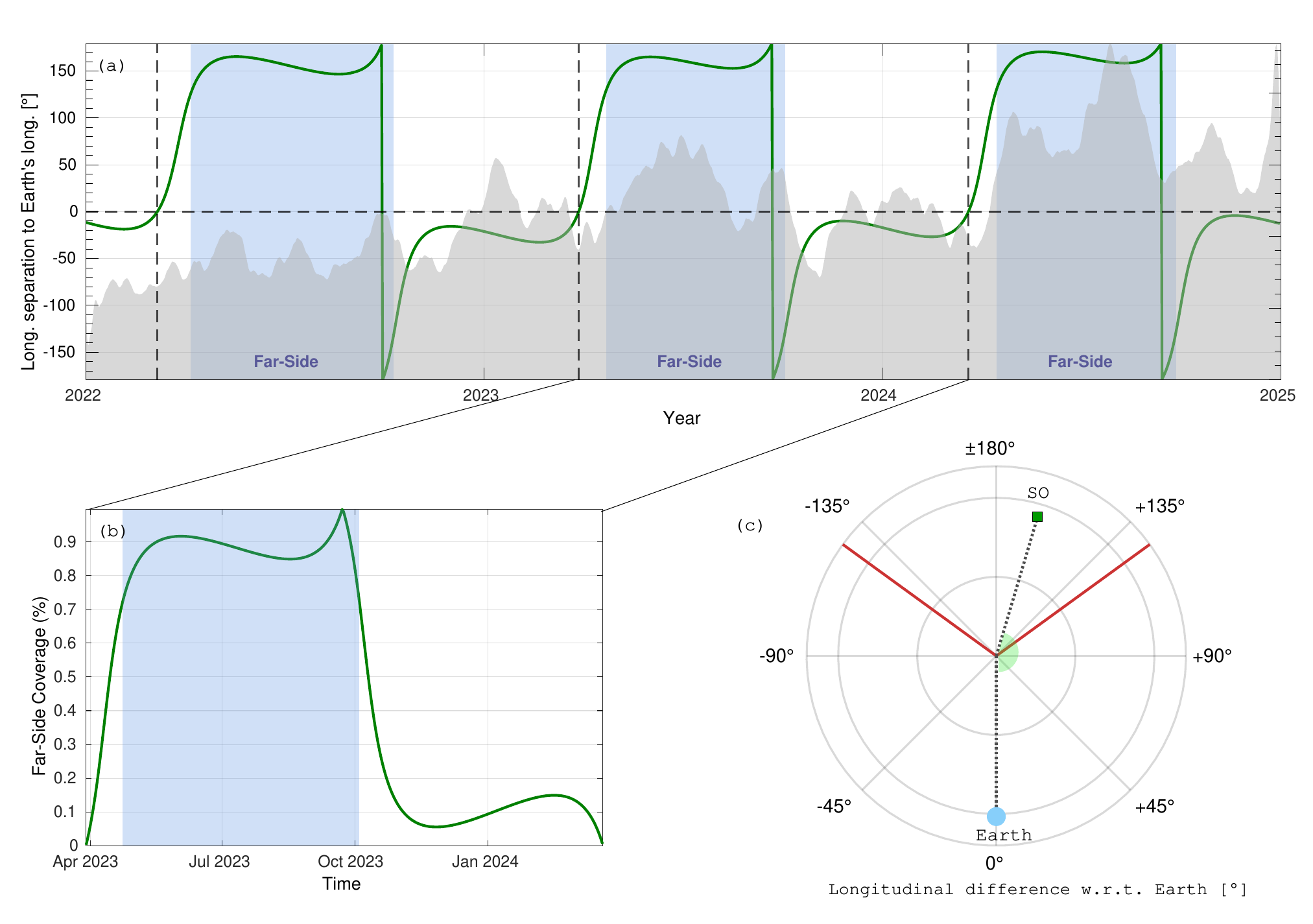} 
    \caption{Solar Orbiter’s far-side dataset coverage, and viewing geometry. (a) Longitudinal separation between Solar Orbiter and Earth through the years 2022 -- 2024, where values approaching $\pm 180^\circ$ signify the predominant far-side vantages. The horizontal dashed line marks the $0^\circ$ separation angle, corresponding to the alignment between SO and Earth in Carrington longitude. The vertical dashed lines indicate the times when Solar Orbiter crosses this alignment. The blue shaded regions represent the periods selected for this study, during which the longitudinal separation exceeded $\pm 126^\circ$, corresponding to the times when Solar Orbiter observes more than two-thirds of the far hemisphere. The gray background area plot shows the smoothed monthly sunspot number for the current Solar Cycle 25 as a reference. (b) Example of the normalized separation angle, scaled relative to the full $180^\circ$ far-side extent, through one of the Solar Orbiter’s orbits, from mid-2023 to early 2024. (c) Polar view of the Solar Orbiter's position (green square) with respect to Earth (pale blue disc at bottom), showing the Stonyhurst longitudinal separation between Earth and Solar Orbiter on 2023 June 19. The green semicircular disk represents the actual angular separation, $\theta$, on that date. The red radial lines indicate the $\pm 126^\circ$ threshold (i.e., 0.7 normalized far-side coverage), used as a cutoff for defining valid far-side intervals in this study.}
    \label{data_range}
\end{figure}

We defined the regions being within a usable far-side visibility, if the normalized angular coverage exceeds a conservative threshold of $x\geq0.7$, which corresponds to observing more than 70\% of the far-side disk (i.e.,$|\theta| \geq 126^\circ$ ). This threshold was chosen to ensure sufficient far-side coverage while avoiding marginal viewing geometries (blue shaded regions in Figure~\ref{data_range}a,b).

Figure \ref{data_range}.a shows the variation of the longitudinal (SO–S–E) separation angle ($\theta$) between Solar Orbiter and Earth during the years 2022 to 2024. The blue shaded regions mark the times when $|\theta|$ exceeded $\pm 126^\circ$, and thus define the intervals used for the analysis as discussed later. The vertical dashed lines correspond to the times when Solar Orbiter crossed the Earth-aligned longitude ($\theta =0$; the horizontal dashed line), while the gray background area provides the solar activity context using the monthly smoothed sunspot number. Panel (b) illustrates an example of the normalized separation angle between two successive crossings for the Earth-aligned longitude ($\theta =0$), from mid-2023 and early 2024, where the blue region denotes the interval with normalized far-side coverage $x$ greater than 0.7. One of the constraints limiting the uniformity of SO/PHI observations during far-side viewing is the asymmetry with respect to $\theta$ in the range $\pm$ 180$^\circ$. During the early years of its mission, Solar Orbiter approached superior conjunction (i.e., +180$^\circ$ separation from Earth) more gradually than it receded. Perihelia typically occurred shortly after superior conjunction, increasing the spacecraft’s radial velocity and thereby limiting the available observation time for the hemisphere rotating into Earth’s view. We filtered for the far-side, and retained the corresponding SO/PHI magnetograms. Finally, panel (c) presents a polar view of the SO-S–E geometry on 19 June 2023. The green semicircular disk shows the actual angular separation on that date, and the red radial lines indicate the $\pm 126^\circ$ threshold used throughout this study. These clearly defined far-side intervals allow a synchronized dataset composed between the helioseismic phase-shift maps, FASTARR-generated AR masks, and SO/PHI magnetograms with minimal temporal offsets and a well-appropriated geometric compatibility.

\subsection*{Spatial Overlap}
\begin{figure}[t]
    \centering
    \includegraphics[width=0.79\textwidth]{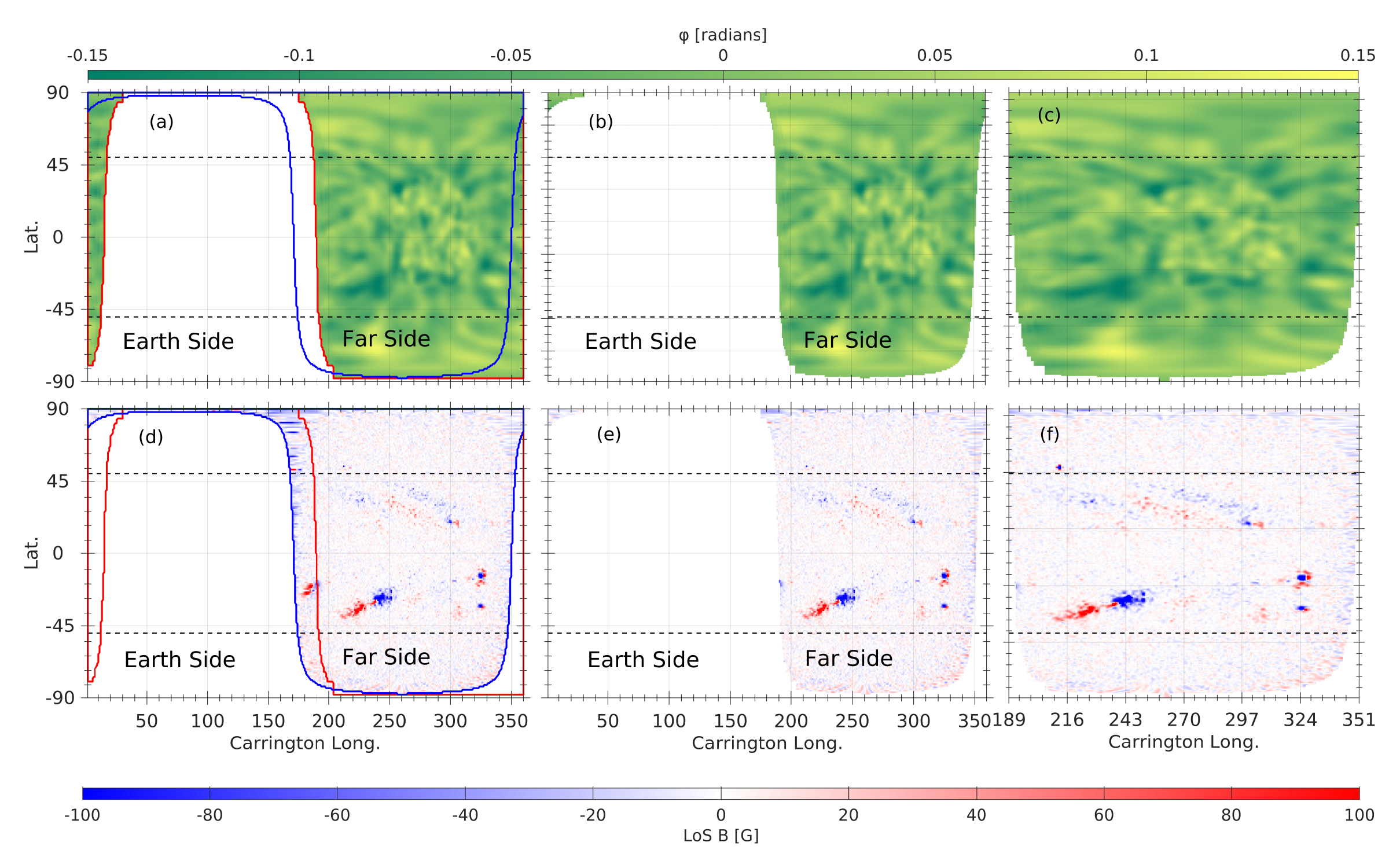} 
    \caption{Comparison between GONG and SO/PHI maps and extraction of overlapping far-side regions on 2022 May 21, 12:00 UT. (a) GONG helioseismic phase-shift map with overlaid visibility boundaries for the full far-side (red outline) and SO/PHI (blue). (b) GONG map masked to retain only the region visible with SO/PHI. (c) Horizontally cropped GONG map. (d) Corresponding SO/PHI magnetogram with visibility boundaries for the far-side in red and SO/PHI in blue. (e) SO/PHI magnetogram truncated to include only the region overlapping with the GONG far-side coverage. (f) Horizontally cropped SO/PHI magnetogram.}
    \label{overlap}
\end{figure}

To compare the far-side GONG phase-shift maps with SO/PHI magnetograms, we first identified regions of overlap between the helioseismic maps and the SO/PHI magnetograms. Figure~\ref{overlap} illustrates a case of favorable viewing geometry. A multi-step masking procedure was applied to isolate the shared visibility domain, enabling direct pixel-level comparisons between helioseismic phase-shift signals and the corresponding LoS magnetic field. The visibility limits were defined in Carrington longitude and heliographic latitude. Figure~\ref{overlap}a,d shows the GONG and SO/PHI maps with their respective visibility boundaries (red and blue contours, respectively). Areas not jointly visible were removed to retain only the common region (Figure~\ref{overlap}b,e), ensuring that comparisons are not biased by differences in viewing geometry.

The truncated maps were geometrically rescaled to fit a uniform square grid in Carrington coordinates. This transformation preserves the local magnetic signal intensity by conserving the total flux within each rescaled pixel, while improving spatial alignment across latitudes. This adjustment facilitates pixel-wise correlation analyses with consistent sampling (Figure~\ref{overlap}c,f) and minimizes systematic biases that could otherwise arise from unequal pixel areas or partial spatial overlap between datasets. These pre-processed datasets are crucial for the next phase, which focuses on detecting far-side ARs and analyzing their magnetic configurations, as described in the next section.

\section*{Methodology}
\label{AR_dataset}

\begin{figure}[t]
    \centering
    \includegraphics[width=0.75\textwidth]{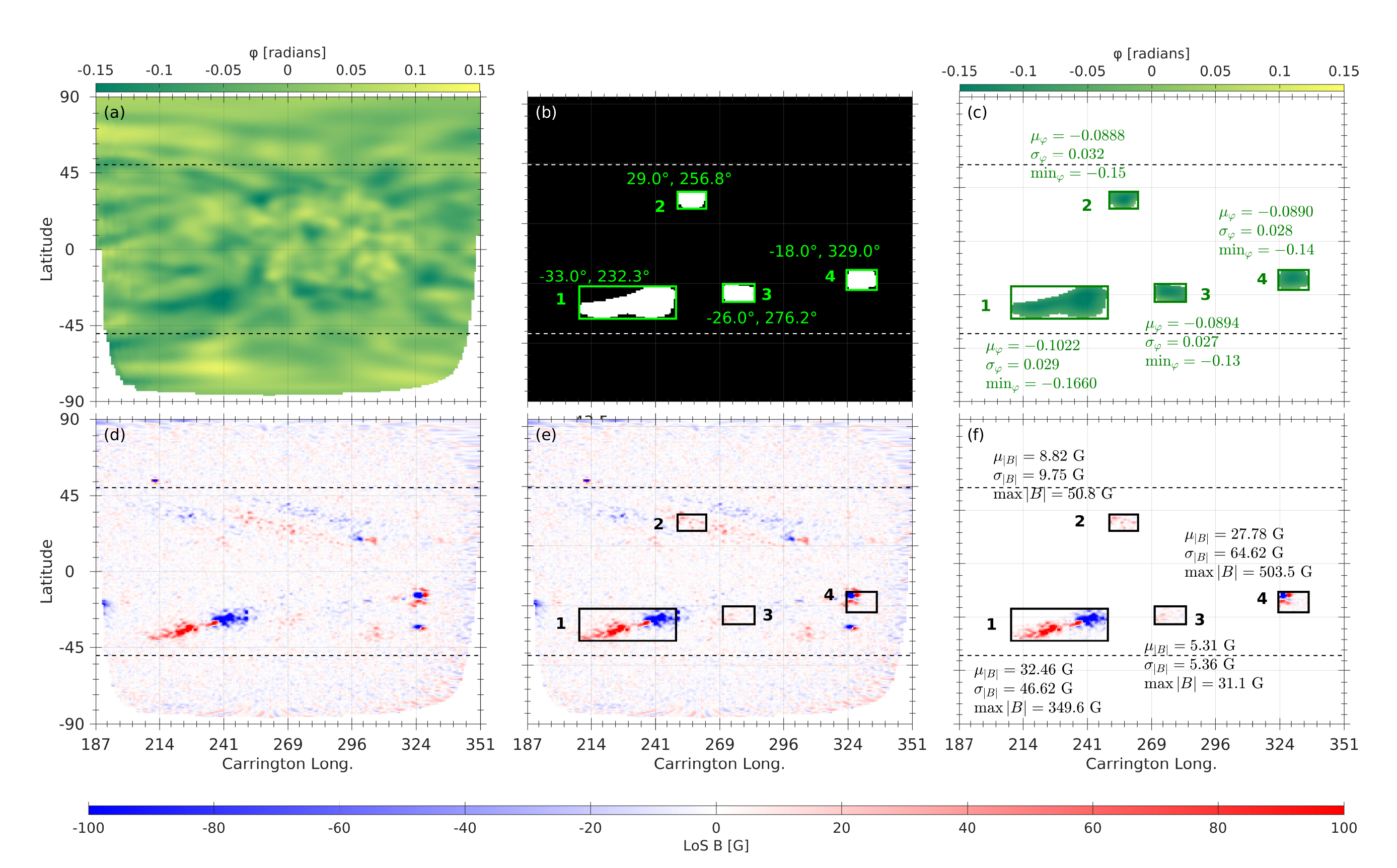} 
    \caption{
    (a) Pre-processed GONG phase-shift map on 2022 May 21, 12:00 UT. (b) Far-side ARs masks, identified using the FASTARR model, overlaid with centroid positions (latitude, and longitude) for each region. (c) Statistical summary of the phase-shift signals $(\varphi)$ within each AR: mean $\mu(\varphi)$, minimum $\min(\varphi)$, and standard deviation $\sigma(\varphi)$.  (d) Co-temporal SO/PHI magnetogram showing the corresponding far-side magnetic field. (e) Same SO/PHI magnetogram with the FASTARR/AR bounded-boxes, overlaid to extract the magnetic characteristics within each identified region. (f) Magnetic property summary for each region: mean of the unsigned LoS magnetic field ($\mu|B|$), standard deviation ($\sigma|B|$), and maximum field (Max$|B|$). 
    }
    \label{AR_data}
\end{figure}

To systematically identify and characterize far-side ARs, we applied our ML model, FASTARR, to the pre-processed GONG phase-shift map (Figure~\ref{AR_data}a), restricted to the overlapping domain with the SO/PHI magnetogram (Figure~\ref{AR_data}d). FASTARR generates a binary mask highlighting candidate ARs (Figure~\ref{AR_data}b), determining their spatial extent. For the example on 2022 May 21, four distinct candidate ARs are detected and labeled. For each, the centroid latitude and Carrington longitude are computed and annotated in the representative binary map. In addition to these geometrical properties, we also compute helioseismic and magnetic characteristics. Figure~\ref{AR_data}c presents statistical properties of the phase-shift signal within each AR, including the mean value $\mu(\varphi)$, standard deviation $\sigma(\varphi)$, and minimum value $\min(\varphi)$, which together characterize the strength and coherence of the helioseismic signal. To assess the magnetic environment of the identified ARs, we extract the corresponding regions from the SO/PHI magnetogram. Figure~\ref{AR_data}d shows the far-side SO/PHI map, while Figure~\ref{AR_data}e overlays the FASTARR-detected AR bounding boxes on the same map, enabling direct spatial comparison between helioseismic detections and the far-side magnetic field distribution. Finally, Figure~\ref{AR_data}f summarizes magnetic properties for each FASTARR-detected AR, including the mean $\mu(|B|)$, standard deviation $\sigma(|B|)$, and  ${\rm Max}(|B|)$ the maximum of unsigned LoS magnetic field.

\begin{figure}[t]
    \centering
    \includegraphics[width=0.85\textwidth]{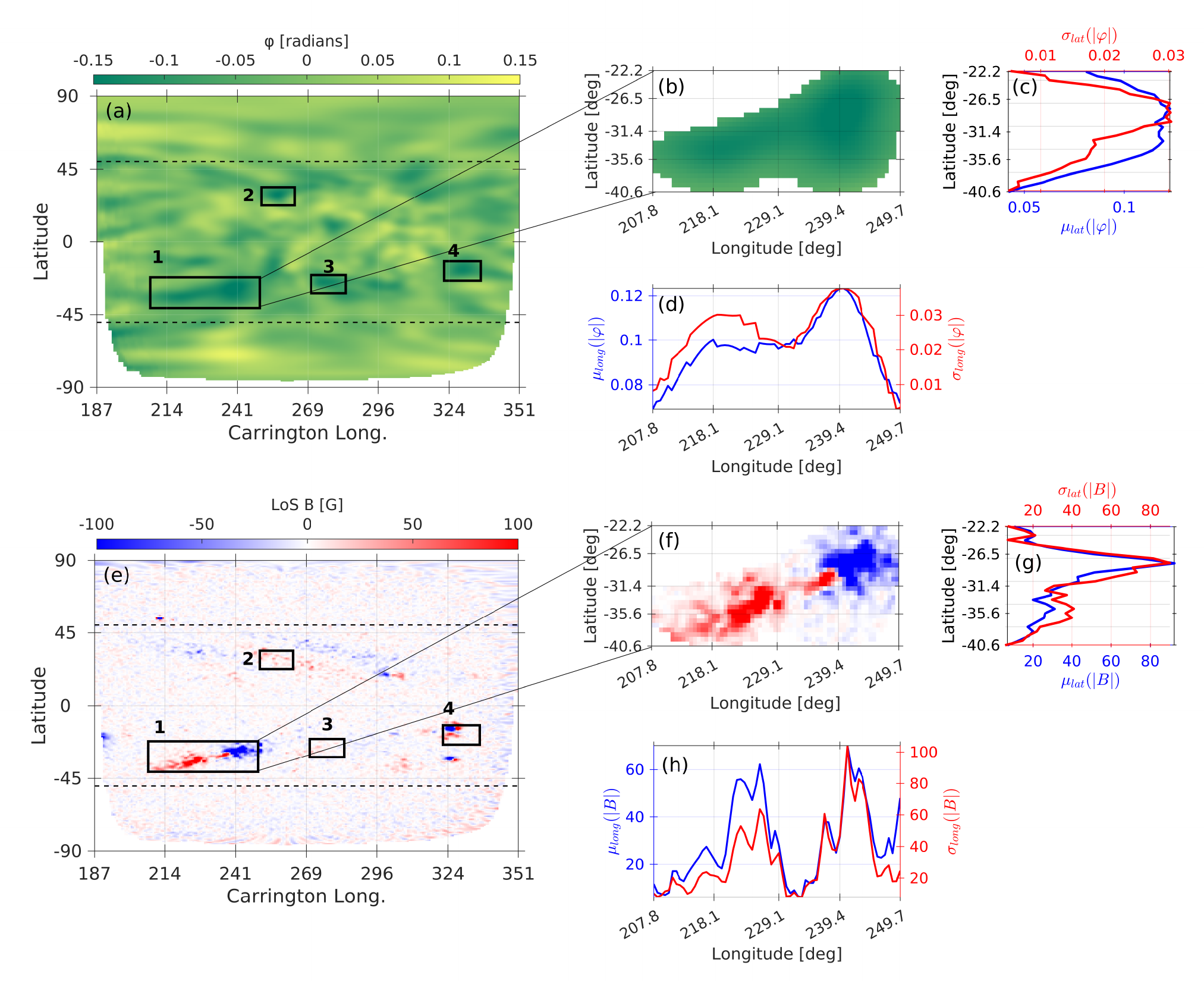} 
    \caption{Comparison of the helioseismic phase-shift and magnetic field properties for the far-side ARs detected. (a) GONG far-side phase-shift map on 2022 May 21 with the identified AR candidates (labeled 1–4). (b) Zoom-in view of the extracted region-1. (c, d) Latitudinal and longitudinal profiles of the mean (blue) and standard deviation (red) of the absolute phase-shift, $|\varphi|$, across the region-1, respectively. (e) Corresponding SO/PHI magnetic field map, co-aligned with the GONG helioseismic map overlaid with the same identified ARs. (f) Zoom in to region 1. (g, h) Latitudinal and longitudinal profiles of the mean (blue) and standard deviation (red) of the unsigned LoS magnetic field $|B|$, respectively.}
    \label{PSvsMAG}
\end{figure}

To assess the consistency between helioseismic phase-shift measurements and direct magnetic observations of far-side ARs, we compared the spatial distribution of phase-shift values with the corresponding SO/PHI LoS magnetograms. Figure~\ref{PSvsMAG} illustrates a representative case for the FASTARR-detected AR~\#1. Panels~\ref{PSvsMAG}b,f show close-up views of the same region based on helioseismic and magnetic data, respectively. Latitudinal and longitudinal cross-sections of the absolute phase-shift signal (Figure~\ref{PSvsMAG}c,d) reveal spatially coherent enhancements in both $\mu_{long}(|\varphi|)$ and $\sigma_{long}(|\varphi|)$, computed as the longitudinal mean and longitudinal standard deviation within the AR boundaries, respectively.

The corresponding distributions of the LoS magnetic field are shown in Figure~\ref{PSvsMAG}g,h. Beyond the spatial agreement, the comparison indicates that enhancements in $\mu_{long}(|\varphi|)$ coincide with regions of strongest unsigned LoS magnetic field, while peaks in $\sigma_{long}(|\varphi|)$ correspond to areas of larger magnetic variability, suggesting that the seismic signal encodes both the intensity and the complexity of the underlying magnetic structure. Together, these results demonstrate a clear qualitative and quantitative correspondence between far-side helioseismic features and magnetic structures, underscoring the potential of integrating both datasets for improved AR monitoring and space-weather forecasting. Regions characterized by more coherent seismic signals and larger mean phase-shift amplitudes tend to coincide with stronger localized magnetic fields, reinforcing the interpretation of acoustic phase delays as reliable proxies for magnetic activity. Importantly, the bipolar structure evident in the helioseismic phase-shift signatures of far-side ARs closely matches the opposite-polarity regions in the SO/PHI magnetograms, providing a practical foundation for allocating their magnetic polarity and enabling a more complete characterization of AR evolution prior to Earth-side visibility.  

\subsection*{AR Selection}
For each identified AR on the far-side helioseismic maps, we used the bi-modal structure of the longitudinal mean and/or variance of the phase-shift values as the main selection criterion. Within each region’s bounding box, longitudinal profiles are derived by computing, for every Carrington longitude column within the AR contour, the mean and standard deviation of the absolute phase-shift values. The longitudinal standard-deviation profiles are then smoothed with a short Gaussian window to suppress pixel-scale noise while preserving the coherent structure. Peaks in the smoothed profiles are detected and ranked by descending amplitude. A region is classified as “bi-modal” if two peaks are present with (i) a separation exceeding 10\% of the profile length and (ii) an amplitude ratio (smaller-to-larger) greater than 0.3. These simple, transparent thresholds provide a reproducible quantification of visibly double-peaked longitudinal structure.

\begin{figure}
    \centering
    \includegraphics[width=0.8\textwidth]{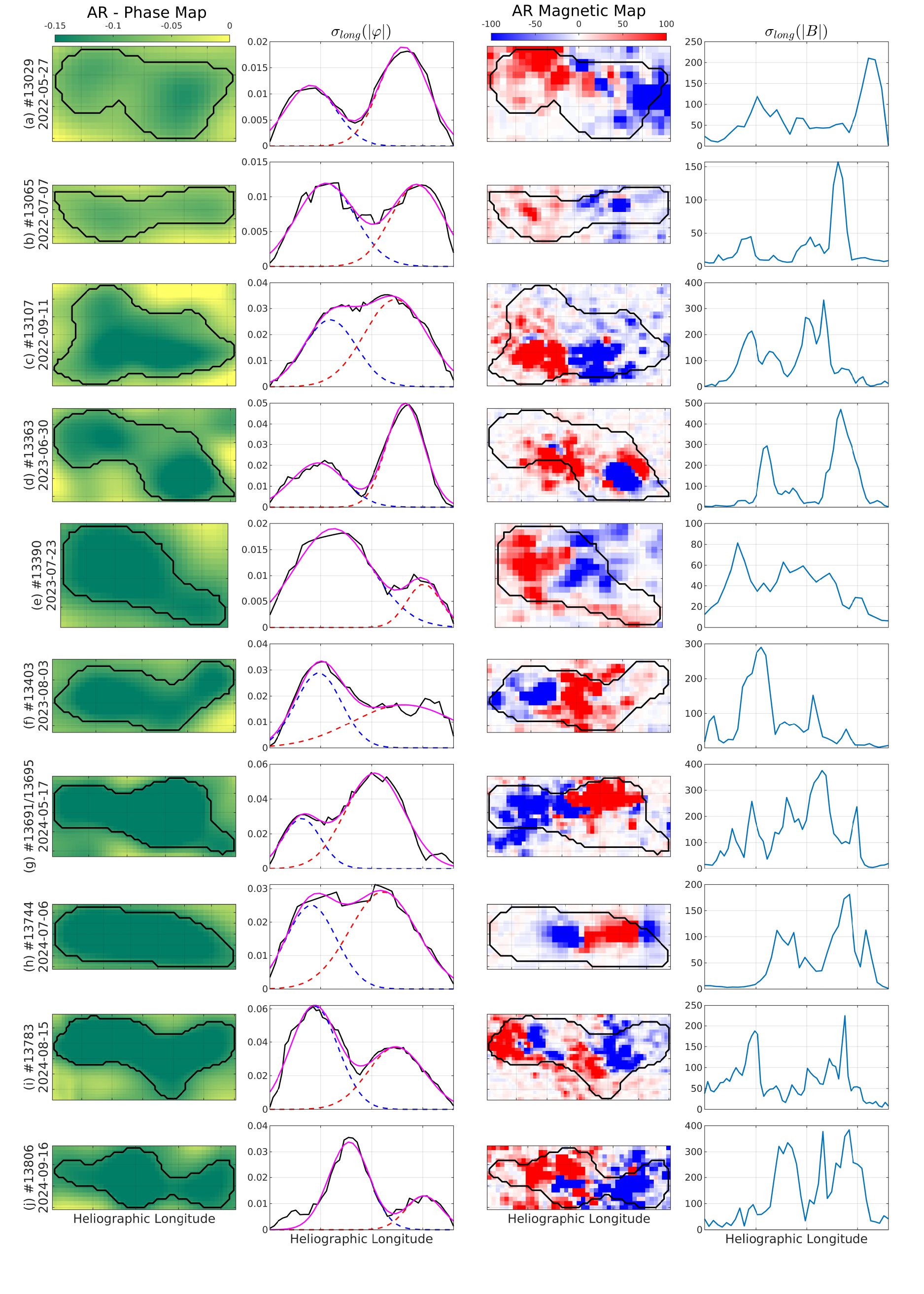} 
    \caption{Comparative analysis of ten far-side ARs using helioseismic phase-shift maps, longitudinal variability, and corresponding magnetic field maps. Each row corresponds to one far-side AR, identified by its NOAA number and far-side crossing date. The first column shows the far-side phase-shift maps with overlaid black contours delineating the AR boundaries automatically identified by the FASTARR model. The second column presents the longitudinal standard deviation of the absolute phase shift, $\sigma_{long}(|\varphi|)$, with the fitted double-Gaussian components (blue and red dashed curves) and their combined profile (magenta line) overlaid on the measured distribution (black line). The third column displays the corresponding far-side magnetic maps with the same FASTARR-derived AR contours overlaid. The fourth column shows the longitudinal standard deviation of the absolute LoS magnetic field, $\sigma_{long}(|B|)$, emphasizing the spatial variability of the magnetic field distribution along the Carrington longitude.}
    \label{AR_select}
\end{figure}

To scale this beyond case studies, we implemented an automated routine that analyzes the entire 2022--2024 dataset, applies the above procedure to every FASTARR-detected region, and retains only those satisfying the bi-modality criteria. This pipeline returned 190 ARs, which constitute the polarity-ready far-side sample used in our subsequent analyses. Figure~\ref{AR_select} displays 10 representative ARs. The first column shows the phase-shift sub-maps, and the second column shows the corresponding longitudinal $\sigma_{long}(|\varphi|)$ profiles, where the dashed curves indicate a two-Gaussian decomposition and the solid curve the reconstructed profile. The third column shows the corresponding SO/PHI LoS magnetogram sub-maps, and the fourth column shows the corresponding longitudinal $\sigma_{long}(|B|)$ profiles. The bi-modality in the ARs helioseismic profiles is co-spatial with corresponding magnetic bipolarity in the magnetograms, reinforcing the interpretation that the phase-shift patterns encode the underlying polarity structure. In well-defined bipolar regions, where the leading and trailing spots are clearly separated, the two-Gaussian peaks align with the magnetic centroids (e.g., Figure~\ref{AR_select}c), and the polarity assignment is reliable and, in practice, near-perfect, consistent with Hale’s law. By contrast, in mixed or complex environments (e.g., when multiple bipoles are longitudinally encased in what is recognized as a single helioseismic signature), the helioseismic signature tends to collapse to two dominant lobes that track only the strongest magnetic concentrations, while weaker or closely spaced polarities are blended (Figure \ref{AR_select}d,h or i). 
This behavior likely reflects the finite spatial resolution of the helioseismic measurements. Therefore, polarity allocation in such cases can be ambiguous and benefits from additional contextual constraints. In its present form, the analysis does not explicitly identify when such compact multi-bipolar configurations underlie a single helioseismic signature, a limitation that naturally motivates future developments incorporating temporal evolution of the seismic signal.

\subsection*{Statistics of $AR(\varphi)$ and $AR(B)$}
To compare helioseismic and magnetic signals on a common footing, we analyze the absolute value of the signed helioseismic phase shift, $|\varphi|$ (hereafter “absolute phase”), and transform the unsigned LoS magnetic field to a logarithmic scale, $\ln|B|$. This logarithmic transform compresses the magnetic field’s dynamic range and stabilizes its variance, yielding a more interpretable, near-Gaussian one–point distribution for AR pixels. We then place both $|\varphi|$ and $\ln|B|$ on a unitless, comparable scale by standardizing to $z$–scores,
\begin{equation}
z \;=\; \frac{x-\bar{x}}{s}\,,
\end{equation}
where $x\in\{\,|\varphi|,\ \ln|B|\,\}$, and $\bar{x}$ and $s$ are the sample mean and standard deviation computed over all AR pixels.

\begin{figure}[H]
    \centering
    \includegraphics[width=0.8\textwidth]{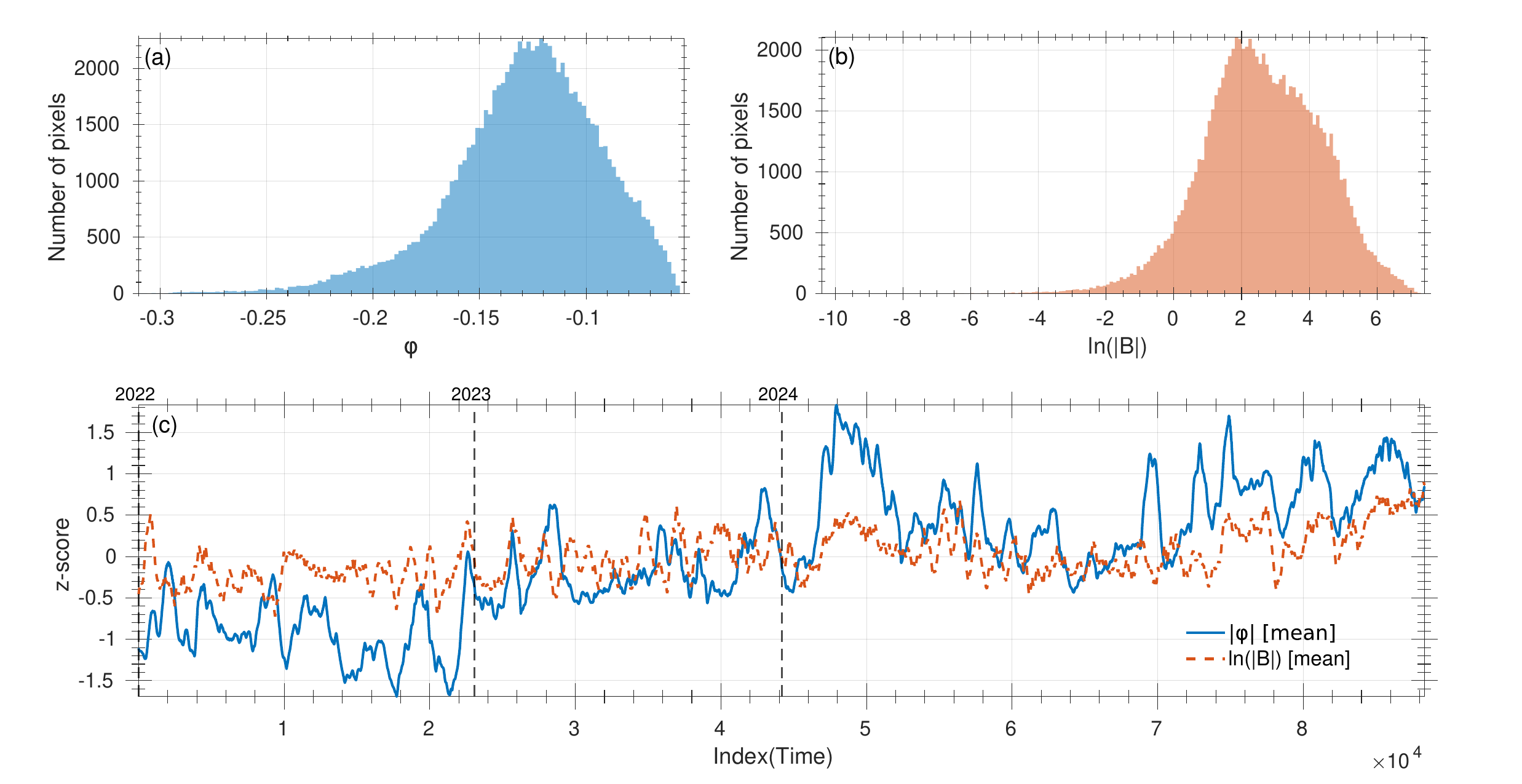} 
    \caption{
    One–point statistics of the phase-shift, $\varphi$, and the magnetic field magnitude, $\ln|B|$, for AR pixels. The histogram plots show the distribution of (a) $\varphi$ and (b) $\ln|B|$ for the selected AR pixels. (c) Means of $|\varphi|$ (solid blue) and $\ln|B|$ (red dashed) versus the sample index, which is used as a proxy for the observational time. The top tick marks annotate calendar years. 
    }
    \label{Hist_all}
\end{figure}

The distribution of the ARs’ helioseismic phase-shift $\varphi$ and magnetic field $\ln|B|$ values (Figure~\ref{Hist_all}a,b) shows distinct, asymmetric profiles that may reflect the underlying physical structures. The helioseismic phase-shift $\varphi$ histogram is entirely shifted toward negative values, with a clear peak near $\varphi \approx -0.125$ and a tail extending toward more negative shifts. This indicates that most AR pixels experience systematic phase delays, consistent with the effect of strong magnetic fields on wave propagation through scattering, absorption, and magnetic canopy interactions. In contrast, the un-signed magnetic field distribution of $\ln|B|$ (Figure~\ref{Hist_all}b) reflects the ARs combined statistics, producing a broad, asymmetric profile that integrates contributions from both weak and strong magnetic environments. The majority of ARs values cluster at moderate field strengths, while the right-extended tail captures the presence of intense magnetic concentrations characteristics across the ensemble. The logarithmic scale allows compressing the wide dynamic range of magnetic intensities, enabling the simultaneous representation of plage, penumbral, and umbral fields within a single unified distribution. Figure~\ref{Hist_all}c shows rolling-mean sequences (600-point moving average) for the absolute helioseismic phase shift, $|\varphi|$, and the unsigned magnetic field, $\ln|B|$. Here, the rolling mean is applied to the sequence of per-pixel $z$–scores after sorting all AR pixels in ascending temporal order according to their associated AR detection time. The 600-point window is used solely as a smoothing operator to highlight long-term trends in the AR population and does not recompute $z$–scores from grouped samples or enforce that the points belong to the same AR. The abscissa represents the samples index and serves as a proxy for selected AR date as all the AR pixels were sorted in a timely ascending order. The two curves exhibit a correlation coefficient of 0.7, with intervals of enhanced magnetic field coincide with an increase in the value of the phase-shift measurements, consistent with the expected association of strong magnetic concentrations with reduced helioseismic signals. Over the span of our AR-dataset through the ascending phase of the current solar cycle 25, from 2022 to 2024, there is a notable increase in the magnetic activity and corresponding values in the phase shift, as the AR complexity rose toward solar maximum.

\subsection*{Phase-Magnetic Calibration}
As the helioseismic imaging detects the far-side ARs by measuring the travel-time or phase-shift perturbations in the acoustic waves propagating through the solar interior, these perturbations arise primarily from local modifications to the effective wave speed. Such modifications are induced by the presence of far-side strong magnetic fields and associated thermal anomalies in the sub-photospheric layers. The dependence of the measured phase shift, $\varphi$, on the magnetic field strength, $|\bf B|$, is inherently non-linear, reflecting the complex physics of wave-magnetic field interactions in a stratified, magnetized plasma. In the weak-field regime (photospheric field strength), the Lorentz force introduces only minor perturbations to the local sound speed. In this limit, the phase-shift response scales approximately with the square of the magnetic field strength, $\varphi \propto$ $|\bf B|^2$, consistent with the second-order sensitivity of acoustic wave propagation to magnetic pressure\cite{Cally_2006}. In the moderate-to-strong field regime, additional physical processes become significant. These include the mode conversion of acoustic waves into magneto-acoustic modes, scattering of p-modes by magnetic inhomogeneities, and partial reflection at magnetically structured boundaries\cite{Cally_1997, Lindsey_2000}. Collectively, these effects reduce the incremental sensitivity of $\varphi$ to further increases in $|\bf B|$, producing a characteristic saturation in the phase-shift response. This saturation generally appears to be well-approximated by a logarithmic-like function of the form,
\begin{equation}
\varphi = \varphi_1 \ln \left( 1 + \frac{|{\bf B}|^2}{B_0^2} \right),
\label{eq_qr_phi_B}
\end{equation}
where $\varphi_1$ is the phase-shift scaling coefficient, $\bf B$ is the magnetic field vector, and $B_0$  expresses a ``magnetic saturation threshold''.\\
 Gonz\'alez Hern\'andez et al. \cite{Gonzalez_2007} recognized this non-linear $\varphi$–$|B|$ relationship in a statistical comparison between GONG far-side helioseismic signatures with Kitt Peak Vacuum Tower (KPVT) and  Synoptic Optical Long-term Investigations of the Sun/Vector Spectromagnetograph (SOLIS/VSM) magnetograms after the corresponding far-side regions rotated into Earth view in which the square, $|B|^2$, of the unsigned LoS magnetic field was approximated as a rough proxy of $|{\bf B}|^2$, hence,
\begin{equation}
 \varphi = \varphi_1 \ln \left( 1 + \frac{B^2}{B_0^2} \right).
\label{irene_eq_qr_phi_B}
\end{equation}

Subsequent analysis by MacDonald et al. \cite{MacDonald_2015}, using near-side helioseismic maps and co-temporal ADAPT magnetograms derived from KPVT  and SOLIS/VSM observations spanning April 2002 to December 2005, yielded an independent phase–magnetic calibration closely matching the non-linear behavior described by Equation~(\ref{eq_qr_phi_B}), thereby reinforcing its validity for far-side applications.

\begin{figure}
    \centering
    \includegraphics[width=0.98\textwidth]{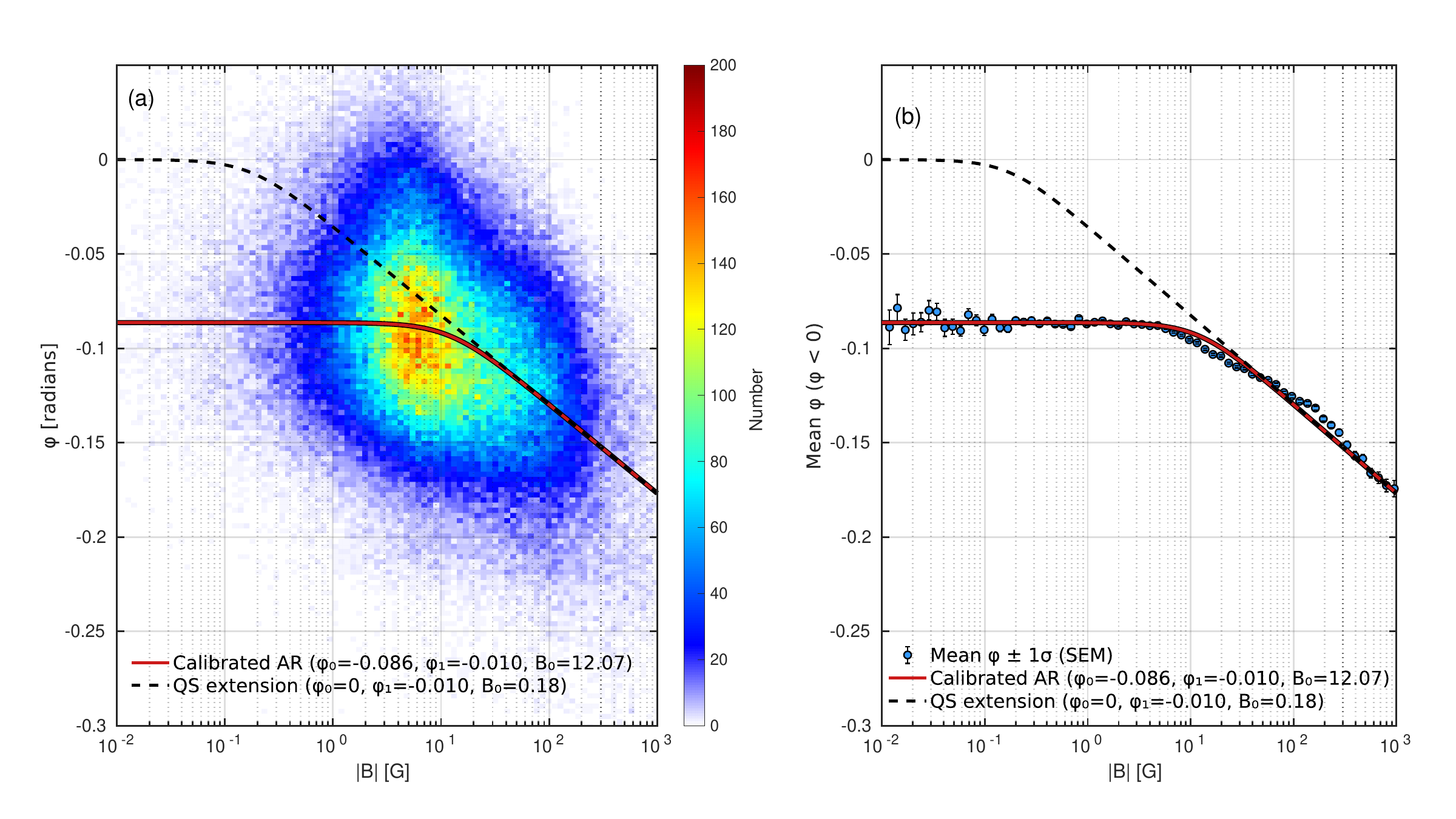} 
    \caption{(a) Two-dimensional histogram of the helioseismic phase shift ($\varphi$) versus the unsigned LoS magnetic field strength ($|B|$) for far-side detections, displayed on a base-10 logarithmic $|B|$ axis. The color scale represents the number of pixels per bin. (b) Mean phase shift values (blue circles) and their associated standard errors ($\pm 1\sigma$, SEM) as a function of $|B|$. In both panels, the solid red line shows the calibrated AR relation derived from the optimized fit ($\varphi_0 = -0.086$, $\varphi_1 = -0.010$, $B_0 = 12.07$~G). The black dashed line corresponds to the quiet Sun (QS) extension
    ($\varphi_0 = 0$, $\varphi_1 = -0.010$, $B_0 = 0.18$~G), which extrapolates the relation toward the quiet-Sun saturation regime.
    }
    \label{phi_mag_fit1}
\end{figure}

Figure~\ref{phi_mag_fit1} shows statistics of GONG helioseismic signatures in the 190 regions recognized by FASTAAR/ARs on the ordinate axis and their LoS magnetic field retrieved by SO/PHI referenced to the abscissa. 
Panel (a) shows the two-dimensional histogram of $(|B|,\varphi)$ displayed as a density map with a logarithmic $|B|$ axis, with the best-fitting model curve over-plotted. The nonlinear relation, i.e., 
\begin{equation}
\varphi = \varphi_0 + \varphi_1 \ln \left( 1 + \frac{B^2}{B_0^2} \right)
\label{irene_eq_qr_phi_B2}
\end{equation}
provides an empirical calibration linking the observed phase shift to the underlying magnetic field in far-side ARs. Panel (b) displays the binned mean phase-shift values (blue circles) together with their uncertainties, shown as vertical error bars corresponding to $\pm1\sigma$ of the standard error of the mean (SEM). The solid red curve shows the optimized calibration relation fitted to the mean values, Equation~\ref{irene_eq_qr_phi_B2}, which captures the empirical AR response of $\varphi$ to $|B|$. The optimized parameters ($\varphi_0 = -0.086$, $\varphi_1 = -0.010$, and $B_0 = 12.07$~G) define the calibrated AR curve, representing the characteristic seismic response of magnetized plasma in the far-side hemisphere. Although the phase–magnetic calibration in Equation~\ref{irene_eq_qr_phi_B2} is expressed using the natural logarithm, the magnetic-field axis in this panel (a) is displayed on a base-10 logarithmic scale for ease of interpretation, where the two representations differ only by a constant scaling factor and do not affect the results.

Following this, the black dashed curve extrapolates the relation toward the quiet-Sun (QS) regime by fixing $\varphi_0=0$ and $\varphi_1=-0.010$ and solving for $B_0$ (here $B_0 = 0.18$\,G) so that the dashed curve matches the calibrated AR relation at the chosen reference field.
In Figure~\ref{phi_mag_fit1}a, these points form prolate blob shaped approximately centered on at the ordinate value of $-0.1$ radian and a magnetic field strength of 10~G. Among the loci prescribed by Equation (\ref{irene_eq_qr_phi_B}), the one that fits these points best is that of the dashed curve.
 
In the narrow context of an optimal fit to just the FASTARR values appearing in Figure~\ref{phi_mag_fit1}, another fits  can be accomplished by using Equation~(\ref{irene_eq_qr_phi_B2}), and allowing $\varphi_0$ to be a free parameter. This fixes the helioseismic signature, $\varphi_0$, at 0.086~radian, far beneath the range, $-0.0035 \pm 0.05$~radian, of ordinate values found in the quiet Sun a safe distance from solar activity. This is made especially attractive by the alternative presentation shown in Figure~\ref{phi_mag_fit1}b, in which the ordinate values in narrow abscissa bins are condensed into the mean value for each bin.

To our present understanding, the discrepancy between the foregoing $\varphi_0$ extrapolation of the quiet Sun and the dashed curve illustrates the fundamental difficulty of non-linearly extrapolating the behavior of a diffuse helioseismic signature from only strongly magnetic regions, to which FASTARR is purposely fashioned to confine itself into weakly magnetic regions a safe distance from dense magnetic field.

In the case of equation~(\ref{irene_eq_qr_phi_B2}), this is not just a matter of weakly magnetic pixels near strong ones between which helioseismic is incapable of resolving.  It is severely contaminated by neutral lines separating strong magnetic poles at which $B$ passes through zero while $\bf B$ is strong and horizontal, a condition we know from observations of ARs in the near hemisphere that elicits {\it strong} helioseismic signatures.
 
\section*{Results}

\subsection*{Far-Side Magnetic Field}
\begin{figure}
    \centering
    \includegraphics[width=\textwidth]{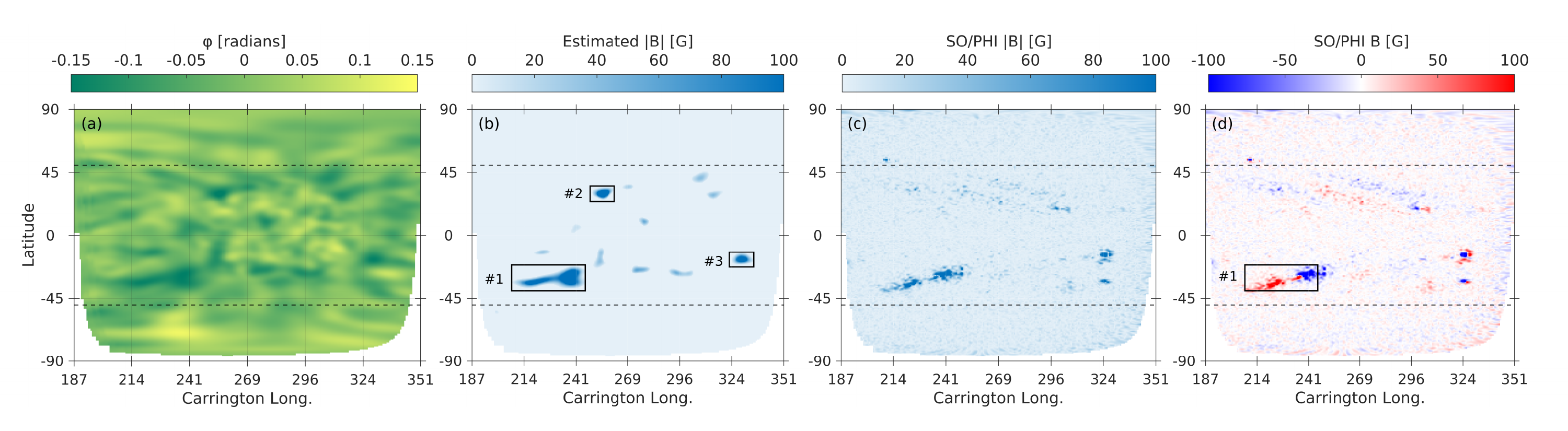} 
    \caption{Reconstruction of far-side signed magnetic field based on helioseismic phase-shift data and validation with SO/PHI. (a) GONG far-side helioseismic phase-shift map ($\varphi$) on 21 May 2022. (b) Estimated unsigned magnetic field strength $|B|$ obtained by applying the calibrated $\varphi$–$|B|$ relation to the phase-shift map, with three AR candidates (\#1–3) highlighted. (c) Co-temporal SO/PHI LoS unsigned magnetic field map for comparison. (d) SO/PHI signed magnetogram ($B$), indicating polarity structure, with the boxed region \#1 carried forward for subsequent comparison.
    }
    \label{phi_mag_fit}
\end{figure}

A central goal of this work is to move beyond the far-side AR identification and demonstrate that helioseismic phase-shift maps can provide quantitative estimates of the magnetic field. Building on the nonlinear calibration between phase shift ($\varphi$) and unsigned LoS magnetic field ($|B|$), we apply an inverse formulation of the calibrated relation (Equation~\ref{irene_eq_qr_phi_B2}) to transform the ARs areas in the helioseismic phase-shift maps into a corresponding magnetic field value. This approach allows us to generate far-side unsigned magnetograms directly from helioseismic data, enabling quantitative comparisons with contemporaneous SO/PHI observations and providing a framework for constructing full-Sun magnetic field maps that extend beyond the Earth-facing hemisphere. The same helioseismic phase-shift map shown previously in Figure~\ref{overlap}, corresponding to 2022 May 21, 12:00 UT, is used here to illustrate this procedure. Panel (a) of Figure~\ref{phi_mag_fit} reproduces the GONG phase-shift map. Applying the nonlinear $\varphi$–$|B|$ calibration, we convert the seismic signatures into an estimated unsigned magnetic field distribution (Figure~\ref{phi_mag_fit}b). Three candidate regions are identified (\#1–3), which are compared against co-temporal SO/PHI observations (panels c–d). The unsigned SO/PHI magnetogram (c) confirms the presence of strong magnetic concentrations at the seismic detection sites, while the signed magnetogram (d) reveals their polarity configuration.

\begin{figure}
    \centering
    \includegraphics[width=\textwidth]{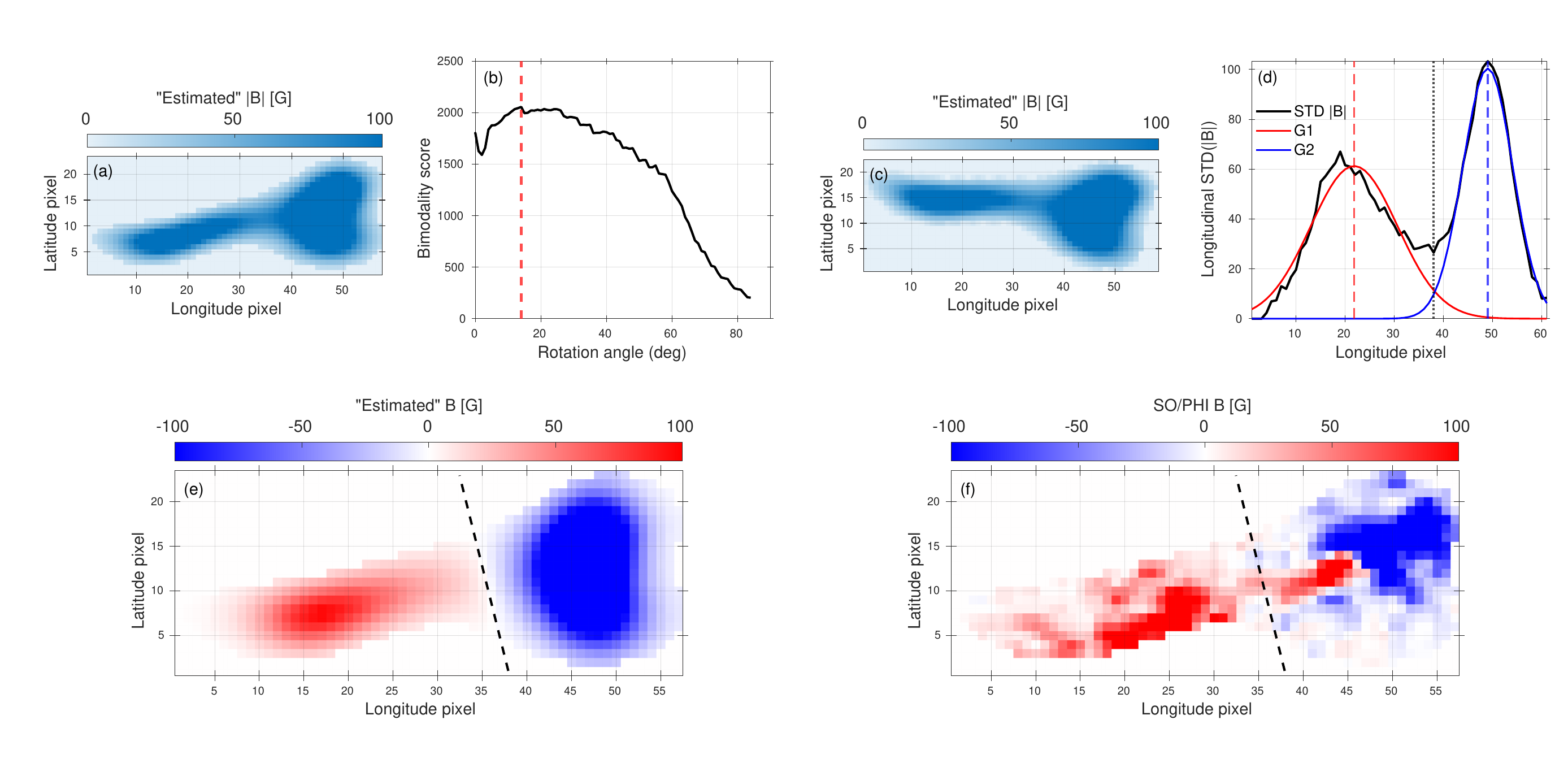} 
    \caption{ Tilt-aware polarity assignment for the representative far-side AR \#1. (a) Estimated unsigned magnetic field intensity, $|B|$, for AR segment \#1  derived from the helioseismic phase-shift map. (b) Bi-modality score as a function of rotation angle, used to identify the optimal orientation that maximizes the separation between the two dominant polarity lobes. The red dashed line marks the optimal rotation angle at $14^\circ$. (c) Same unsigned magnetic field intensity for AR segment \#1 as in panel (a) after rotation by the optimal angle $(14^\circ)$, aligning the dominant bipolar structure along the analysis axis. (d) Longitudinal standard deviation profile of the rotated AR segment \#1 (black), together with the two-Gaussian decomposition, where $G_1$ (red) and $G_2$ (blue) represent the two dominant polarity components. (e) Reconstructed signed magnetic field using the optimized polarity inversion line (dashed black line) derived from the tilt-aware bi-modality analysis. (f) Co-temporal SO/PHI signed LoS magnetogram for the same region, shown for comparison.}
    \label{phi_mag_tilt}
\end{figure}

We developed a polarity-assignment framework that leverages the longitudinal variance of the estimated $|B|$ within the identified AR bounding boxes to infer its magnetic polarity configuration. The approach is motivated by the fact that strong far-side ARs typically exhibit two dominant magnetic concentrations, which manifest as enhanced longitudinal variability in the inferred unsigned magnetic field. For each selected AR, the unsigned magnetic field map is first analyzed within its bounding box (Figure~\ref{phi_mag_tilt}a). The map is then systematically rotated by a trial angle $\theta$, and for each orientation the longitudinal profile of the standard deviation of the unsigned magnetic field is computed as

\begin{equation}
\sigma_\theta(x) = \mathrm{STD}_{y}\left(|B|_\theta\right),
\end{equation}

where $|B|_\theta$ denotes the rotated $|B|$ map and the standard deviation is evaluated along the latitudinal direction. To quantify the degree of bipolar separation as a function of rotation angle, we define a bi-modality score $S(\theta)$ based on the structure of $\sigma_\theta(x)$. For each trial angle $(\theta)$, the two most prominent peaks in the $\mathrm{STD}_{y}(|B|_\theta)$ profile are identified, with peak locations $x_1$ and $x_2$ and corresponding amplitudes $P_1$ and $P_2$. The bi-modality score is then defined as

\begin{equation}
S(\theta) = |x_2 - x_1| \times \min(P_1, P_2),
\end{equation}

which combines the spatial separation of the two dominant magnetic lobes with a measure of their relative prominence. This definition favors the configurations in which the two spatially separated magnetic concentrations of comparable strength are present, while suppressing those cases dominated by a single peak or noise-induced structure (Figure~\ref{phi_mag_tilt}b). The rotation angle that maximizes $S(\theta)$ (vertical dashed line in Figure~\ref{phi_mag_tilt}b) is adopted as the optimal reference frame for the polarity separation, effectively aligning the analysis axis with the AR's intrinsic tilt.

After rotation to the optimal angle (Figure~\ref{phi_mag_tilt}c), the longitudinal standard-deviation profile exhibits a pronounced bi-modal structure, reflecting the presence of two dominant magnetic concentrations. This profile is modeled as the sum of two Gaussian components, $G_1(x)$ and $G_2(x)$, which represent the large-scale magnetic lobes associated with the AR's leading and trailing polarities (Figure~\ref{phi_mag_tilt}d). The separation between the two components characterizes the bipolar structure, while their intersection point defines the most probable location of the polarity inversion line (PIL), providing an objective and data-driven estimate of the large-scale polarity boundary. To avoid enforcing a hard binary polarity assignment, a continuous polarity-weighting function $w(x)$ is constructed from the normalized difference of the two fitted Gaussian components,

\begin{equation} 
w(x) = \frac{G_1(x) - G_2(x)}{G_1(x) + G_2(x)} . 
\end{equation} 

This weighting is further modulated by the local longitudinal variability, quantified by the standard-deviation profile $\sigma(x)$,

\begin{equation} 
w'(x) = w(x) \frac{\sigma(x)}{\max[\sigma(x)]} , 
\end{equation} 

thereby assigning higher confidence to regions of strong magnetic variability and suppressing noise in weak or diffuse areas. The final signed magnetic field is then obtained as

\begin{equation}
    B'(x,y) = |B(x,y)| \, w'(x) \, s_{\mathrm{pol}} \, s_{\mathrm{hem}} \,,
\end{equation}

where $s_{\mathrm{pol}}$ denotes the local polarity sign determined by association with the Gaussian components ($s_{\mathrm{pol}}=+1$ for pixels associated with $G_1$ and $s_{\mathrm{pol}}=-1$ for those associated with $G_2$), and $s_{\mathrm{hem}}$ is a hemispheric sign factor imposed by Hale’s polarity law. According to Hale’s law, the leading and trailing polarities exhibit opposite configurations in the northern and southern hemispheres and reverse from one solar cycle to the next \cite{2020_Paul}. For SC~25, the leading polarity is positive and the trailing polarity is negative in the northern hemisphere, with the opposite configuration in the southern hemisphere. Applying this framework yields, and after rotation back to the optimal tilt angle, a smoothly varying, polarity-resolved magnetic field distribution that respects both the inferred PIL geometry and the magnetic polarity pattern (Figure~\ref{phi_mag_tilt}e). The reconstructed signed magnetic field for AR~\#1 shows a coherent bipolar structure whose orientation and polarity separation closely match the co-temporal SO/PHI signed magnetogram (Figure~\ref{phi_mag_tilt}f), demonstrating that the rotation-optimized bi-modality analysis captures the dominant magnetic organization of the identified far-side AR. These steps enable for the first time the construction of polarity-resolved far-side magnetograms based on helioseismic observations.

This case study provides a clear validation that helioseismic phase-shift maps, when combined with Gaussian decomposition, Hale’s law constraints, and empirical calibration, encode sufficient information to reconstruct both the magnitude and polarity of far-side AR magnetic fields. In contrast to earlier far-side techniques that yielded only unsigned proxies, this framework enables the generation of polarity-resolved magnetograms. Such a capability is critical for constructing global full-Sun magnetic field maps, constraining coronal extrapolations, and improving space-weather modeling of the solar wind and CME background fields.

A quantitative comparison between the estimated signed magnetic field and the co-temporal SO/PHI observations was performed for AR segment \#1 (Figure \ref{phi_mag_tilt}e-f, respectively). For the signed magnetic field, the comparison yields a mean absolute error of MAE = 26.99~G and a root mean square error of RMSE = 57.82~G. The linear Pearson correlation coefficient is $r = 0.465$, while the Spearman rank correlation coefficient reaches $\rho = 0.718$, indicating that the relative ordering and spatial structure of the magnetic field are well preserved. Using a threshold of $|B| \geq 20$~G, the recovered polarity achieves an accuracy of 90.8\%. For the unsigned magnetic field, the agreement improves, with MAE = 23.61~G and RMSE = 48.44~G. The Pearson correlation coefficient between the estimated and observed $|B|$ values is $r = 0.528$, demonstrating a moderate but statistically significant correspondence in magnetic field strength.

\subsection*{Far-Side Transit of ARs13664/13668 and AR13663}
We present a case study for the far-side transit of ARs 13664/13668 and AR 13663 during May 2024. The sequence of extreme solar activity in May 2024\cite{2025ApJ...979...49H}, culminating in the geomagnetic disturbances known as the "Mother’s Day Storm" or "Gannon Storm," was associated with the emergence and interaction of several large ARs in the southern hemisphere. AR 13663 was the first to appear in the northern hemisphere, contributing significantly to the early flaring phase, with multiple M- and X-class\cite{2025ApJ...987..134L}. Shortly thereafter, AR 13664 emerged in the southern hemisphere, rapidly becoming the dominant source of eruptive activity. This region was responsible for the series of Earth-directed CMEs that drove the most severe geomagnetic disturbances of the event. The later emergence of AR 13668, also in the southern hemisphere and adjacent to AR 13664, added further magnetic complexity and contributed to the extended flare productivity of the AR complex.

\begin{figure} 
    \centering
    \includegraphics[width=0.9\textwidth]{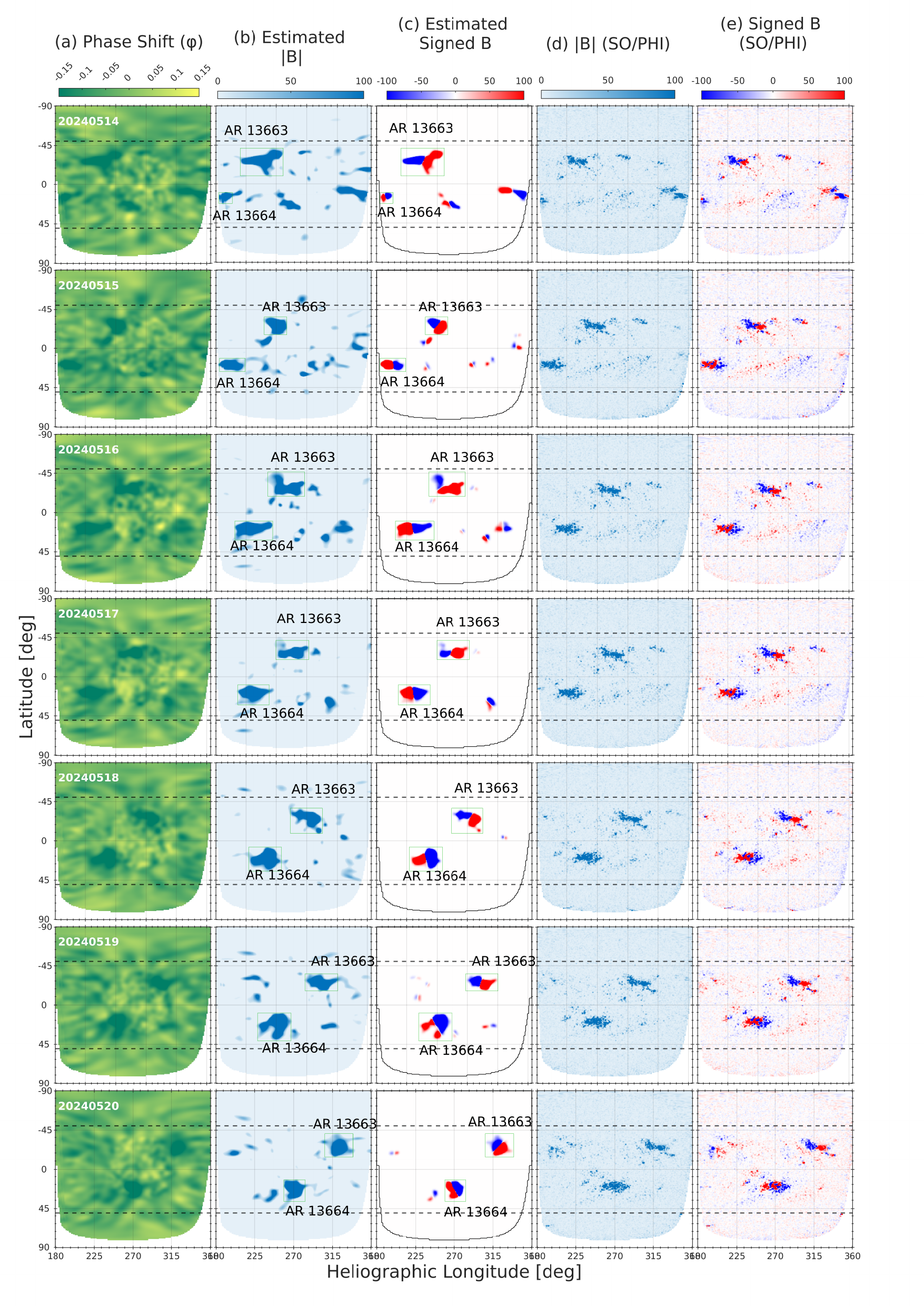} 
    \caption{Far-side transit of NOAA AR 13663 and ARs 13664/13668 during 14 -- 20 May 2024, analyzed using helioseismic phase-shift maps from GONG and contemporaneous magnetograms from SO/PHI. Each row corresponds to a daily observation/measurement across the far-side passage. (a) Helioseismic phase-shift map ($\varphi$) highlighting the ARs acoustic signatures. (b) Estimated unsigned magnetic field $|B|$ derived from the inverted $\varphi$–$|B|$ calibration. (c) Reconstructed signed magnetic field maps obtained through the Gaussian decomposition of the longitudinal variance and polarity ordering via Hale’s law. (d) Co-temporal SO/PHI unsigned LoS magnetograms, and (e) SO/PHI signed magnetograms. 
    }
    \label{Case_01}
\end{figure}

\begin{figure} 
    \centering
    \includegraphics[width=0.9\textwidth]{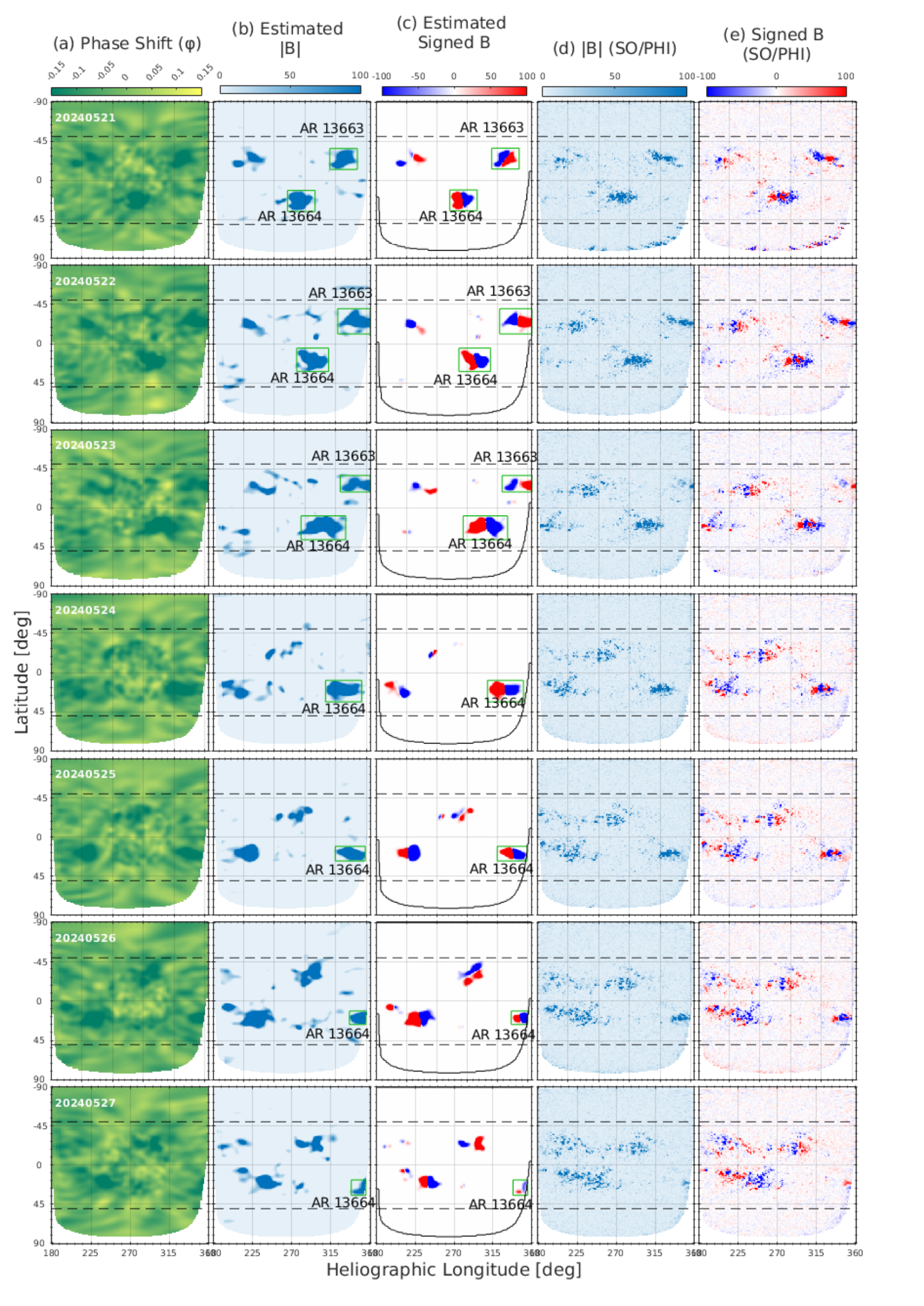} 
    \caption{Same as Figure~\ref{Case_01} but corresponding to the period 21 -- 27 May 2024. 
    }
    \label{Case_02}
\end{figure}

We examined the far-side transit of ARs 13664/13668 and AR 13663 using a combined dataset of GONG helioseismic phase-shift maps and the estimated un-signed/signed maps, together with the temporally matched SO/PHI magnetograms. Figures~\ref{Case_01} and  \ref{Case_02} show the evolution of AR 13663 and AR 13664/13668 during their far-side passage, spanning 14–27 May. Both ARs appear prominently in the phase-shift maps (Column a), with AR 13663 dominating the early phase and ARs 13664/13668 rapidly becoming the main contributor to far-side magnetic activity. The helioseismic-based estimated unsigned and signed magnetic maps (columns b–c) reproduce the morphology and strength of the co-temporal SO/PHI observations (columns d–e), validating the ability of helioseismic phase-shift data to provide quantitative magnetic field estimates on the solar far-side. In the case of ARs 13664/13668, the strong and coherent phase-shift signatures maintained throughout the transit confirmed the survival of a magnetically complex system beyond the west limb. Complementary SO/PHI magnetograms provided critical validation of these seismic inferences. 

Whereas the helioseismic maps captured the overall presence and coherence of ARs 13664/13668, SO/PHI offered direct measurements of the evolving magnetic distribution and polarity structure from a vantage away from the Sun–Earth line. These magnetograms reveal that while substantial magnetic field persisted during the far-side passage, the region underwent clear fragmentation and weakening compared to its initial near-side emergence. Such detailed magnetic information is inaccessible to seismic imaging alone, underscoring the necessity of combined approaches.

The reappearance of AR 13664 as AR 13697 on 27 May 2024 illustrates the progressive dispersal of magnetic field and reduction in non-potentiality following its major eruptive phase. This decay is consistent with the broader behavior of super ARs by exhibiting a rapid energization during their emergence followed by magnetic fragmentation and a declining in flare productivity after their most eruptive phases\cite{Zirin_1987,Toriumi_2019}. Nevertheless, AR 13697 remained flare-productive, highlighting that even decaying regions retain sufficient free energy to power significant eruptions over multiple rotations.

By validating the helioseismic-based magnetic maps against direct magnetic observations, we demonstrate the capacity of far-side helioseismology not only to track the survival of such super ARs, but also to provide critical inputs for space-weather forecasting during intervals when direct far side observations are unavailable. The results highlight the complementary diagnostic power of these datasets in tracking large-scale AR evolution when the regions are beyond Earth’s direct view.

The case of ARs 13664/13668 underscores the scientific and operational value of combining GONG far-side helioseismology with SO/PHI magnetography. Together with the Earth-side observations, these datasets provide a comprehensive view of magnetic persistence, magnetic evolution, and eruptive potential across a far-side transit. Such coordinated approaches are essential for advancing both our physical understanding of super AR and the operational forecasting of extreme space-weather events.
 
\section*{Discussion}
The foregoing results secure that helioseismic monitoring of the Sun's far hemisphere, together with other resources from our present understanding of the general disposition and evolution of magnetic regions, can give us considerably more insight into the prospective magnetic disposition of an AR in the Sun's far hemisphere than just helioseismic maps alone. A calibrated, non-linear relationship between helioseismic phase shift ($\varphi$) and the corresponding unsigned LoS magnetic field ($|B|$) enables us to convert GONG phase-shift maps into unsigned magnetograms. Coupled with an objective longitudinal bi-modality diagnostic and Hale’s hemispheric polarity rule,  this approach produces polarity-resolved far-side magnetograms that align well with co-temporal SO/PHI observations from 2022–2024. Case studies --  such as  the May 2024 complex involving AR~13663 and ARs~13664/13668 --  demonstrate consistent morphological and polarity agreement throughout far-side transit and reappearance. Ensemble statistics from ~190 ARs meeting our selection criteria further confirm that the calibrated
  $\varphi$--$\ln{|B|}$ relation is stable over a wide range of magnetic conditions and phases of Solar Cycle~25.

Physically, the inferred magnetic scale factor ($B_{0} \approx 12.07$~G) is consistent with expectations from wave--magnetic interactions in stratified, magnetized plasma. In weak to moderate fields, increased Lorentz forces and magnetic pressure perturb the effective wave speed and path, producing phase delays that scale approximately with $B^{2}$. At higher magnetic fields, mode conversion, scattering by inhomogeneities, and partial reflection at magnetic canopies reduce incremental sensitivity, producing the observed logarithmic-like saturation. This calibration is not an artifact of the few isolated cases; rather, it reflects a reproducible property of the far-side seismic signal, consistently observed across hundreds of regions, multiple epochs, and diverse viewing geometries.

From an operational perspective, our results help bridge a long-standing gap in full-Sun magnetic monitoring. Although near-side magnetograms are critical for assimilating data, modeling solar wind, and extrapolating coronal fields,  the lack of direct far-side magnetograms has historically compelled forecasters to depend on proxies with limited magnetic insight. Our goal is to produce daily, low-latency, polarity-resolved far-side maps at 6-hour cadence from an existing, stable ground-based network. When integrated with near-side magnetograms (e.g., in synoptic or synchronic frameworks), this enables construction of 360$^{\circ}$ magnetic boundary conditions for coronal and heliospheric models, providing a critical input toward improved solar-wind prediction frameworks, and CME environment modeling. The May 2024 sequence underscores the value of such continuity, where AR~13664's near-side emergence heralded a hyper-active epoch, its far-side evolution and fragmentation before reappearance could be tracked and quantified in the absence of direct Earth-view magnetograms.

Methodologically, our work incorporates three advances that, together, make polarity-resolved mapping feasible. First, FASTARR provides consistent, automated AR masks on the far side, enabling region-wise statistics and robust handling of diffuse peripheries. Second, the global $\varphi$--$|B|$ calibration, estimated with an empirical scan of the magnetic  parameter $B_{0}$ and joint fits for $(\varphi_{0}, \varphi_{1})$, as illustrated by the fitting behavior in Figure 7, yields a compact parametric bridge from seismic to magnetic domains with interpretable uncertainty structure. Third, the polarity inference relies on a measurable feature of the seismic (or inferred $|B|$) field, the longitudinal standard-deviation bi-modality within AR bounds, so that polarity boundaries are tied to the underlying signal morphology rather than heuristic geometry alone. The resulting polarity separation, structured according to Hale’s law, accurately reproduces leading -- trailing patterns across both hemispheres during Cycle~25 and, importantly, aligns with contemporaneous SO/PHI polarity maps even during complex, evolving AR passages. Small but systematic centroid offsets ($\sim$1$^{\circ}$) between helioseismic and magnetic peaks likely reflect differences in sensitivity, height of formation (helioseismic sensitivity to sub-photospheric structure vs.\ photospheric LoS field), and residual geometrical distortions. Nevertheless, the offsets remain small compared to AR dimensions and do not affect polarity classification.\\

Although the approach has proven effective, its performance is inherently constrained in cases where helioseismic signatures deviate from clear, single-bipole morphologies. At present, our selection favors ARs exhibiting pronounced longitudinal bi-modality. The very compact, multi-polar, or $\delta$-configuration regions with tangled polarity may produce weaker or multiple peaks, increasing ambiguity. Reliance on Hale’s law is appropriate on average; however, anti-Hale ARs, whose leading polarity violates the cycle-dependent hemispheric ordering, are not explicitly identified in the current implementation and may therefore receive incorrect polarity assignment. Also, incorporating latitudinal Joy’s-law tilts, temporal continuity across successive maps, and data-driven priors on bi-pole orientation should reduce mis-assignment risk. Our calibration uses LoS magnetograms from SO/PHI. To support assimilation into radial field models, applying systematic corrections and cross-calibrating with GONG or other front-side LoS magnetograms will enhance consistency. At high latitudes and near the far-limb boundaries of the seismic aperture, foreshortening and filtering reduce signal-to-noise (S/N) ratio. Thus, uncertainty maps that propagate phase noise, duty-cycle thresholds, and local fit residuals into per-pixel $|B|$ and polarity confidence will help users weight the product in assimilation. Finally, although the 18-hour temporal averaging enhances S/N, it can smear rapid emergence phases, where adaptive windows tied to data quality and regional dynamics could preserve transients without sacrificing robustness.\\

Despite all of these caveats, the demonstrated agreement with SO/PHI LoS magnetograms across diverse geometries provides an empirical validation that far-side helioseismology can support quantitative, polarity-aware magnetic mapping suitable for research and operations. We envision three concrete extensions. First, integrating EUV intensity and coronal-hole context can regularize polarity inference in complex regions by adding complementary constraints on connectivity and open-flux distribution. Second, learning a joint latent space linking $(\varphi, |B|, \mathrm{EUV})$ could replace the analytic calibration with a hybrid physics-informed model that outputs magnetic field signal intensities, polarity, and uncertainties end-to-end while remaining interpretable. Third, systematic impact studies by running paired forecast cycles with and without far-side signed maps can quantify the downstream gains in solar wind speed, polarity sector structure, and CME background fields. Together, these steps will move far-side helioseismology from qualitative detection toward a fully assimilable magnetic boundary condition for global heliophysics, enabling more resilient forecasting during intervals when Earth-view magnetograms are blind.

\section*{Funding}
KJ was partially supported by the NASA DRIVE Center award 80NSSC20K0602 to Stanford University. The German contribution to SO/PHI is funded by the BMWi through DLR and by MPG central funds. The Spanish contribution is funded by AEI/MCIN/10.13039/501100011033/ (RTI2018-096886-C5, PID2021-125325OB-C5) and ERDF "A way of making Europe"; "Center of Excellence Severo Ochoa" awarded to IAA-CSIC (SEV-2017-0709, CEX2021-001131-S). The French contribution is funded by CNES.

\bibliography{sample}

\section*{Acknowledgments }

 AH and KJ were partially supported by the NSF Windows of the Universe - Multi-Messenger Astrophysics (WoU-MMA) grant  to the National Solar Observatory. KJ also acknowledges partial support from NASA-DRIVE Center award 80NSSC20K0602  to Stanford. This work utilizes GONG data obtained by the NSO Integrated Synoptic Program, managed by the National Solar Observatory, which is operated by the Association of Universities for Research in Astronomy (AURA), Inc. under a cooperative agreement with the National Science Foundation and with a contribution from the National Oceanic and Atmospheric Administration. The GONG network of instruments is hosted by the Big Bear Solar Observatory, High Altitude Observatory, Learmonth Solar Observatory, Udaipur Solar Observatory, Instituto de Astrof\'{\i}sica de Canarias, and Cerro Tololo Inter-American Observatory. Solar Orbiter is a space mission of international collaboration between ESA and NASA, operated by ESA. We are grateful to the ESA SOC and MOC teams for their support. The German contribution to SO/PHI is funded by the BMWi through DLR and by MPG central funds. The Spanish contribution is funded by AEI/MCIN/10.13039/ 501100011033/ (RTI2018-096886- C5, PID2021-125325OB-C5) and ERDF "A way of making Europe"; "Center of Excellence Severo Ochoa" awarded to IAA-CSIC (SEV-2017-0709, CEX2021- 001131-S). The French contribution is funded by CNES. 

\section*{Author contributions statement}
K.J. designed the project and generated far-side helioseismic maps. H.S. and D.O.S calibrated SO/PHI data, selected the far side SO/PHI LoS magnetograms and remapped them to Carrington coordinates. A.H. developed the methodology, conducted the analysis and drafted the manuscript. A.H., K.J., and C.L. discussed the results and contributed to the manuscript. All authors reviewed and approved the final manuscript. 

\section*{Additional information}
\textbf{Competing interests}

The authors declare no competing interests.
\end{document}